\documentclass[apl, twocolumn,tightenlines,superscriptaddress]{revtex4-2}

\usepackage[UKenglish]{babel}
\usepackage{amsthm}
\usepackage{amsmath}
\usepackage{bbm, dsfont}
\usepackage{graphicx}
\usepackage[usenames,dvipsnames]{color}
\usepackage[colorlinks=true,citecolor=magenta,urlcolor=blue]{hyperref}
\usepackage[table]{xcolor}
\usepackage{colortbl}
\usepackage{color}
\usepackage{multirow}
\usepackage{hhline}
\usepackage{epstopdf}
\usepackage{booktabs}
\usepackage[normalem]{ulem}
\usepackage{siunitx}
\usepackage{enumitem}
\usepackage{braket}
\usepackage{siunitx}
\usepackage{textcomp}
\usepackage{booktabs}
\usepackage[textwidth=3cm, textsize=footnotesize]{todonotes}
\usepackage[font=small,skip=3pt]{caption}
\usepackage{bbold}
\usepackage[percent]{overpic}
\usepackage{makecell}
\usepackage{multirow}
\usepackage{float}
\usepackage{textcomp}

\newcommand{\affiliationUVigo}{Vigo Quantum Communication Center, University of Vigo, Vigo E-36310, Spain}
\newcommand{\affiliationComGroup}{Escuela de Ingeniería de Telecomunicación, Department of Signal Theory and Communications, University of Vigo, Vigo E-36310, Spain}
\newcommand{\affiliationAtlanttic}{AtlanTTic Research Center, University of Vigo, Vigo E-36310, Spain.}
\newcommand{\affiliationGeneva}{Université de Genève, Department of Applied Physics, Quantum Technologies, Rue de l'Ecole-De-Médecine 20 1205 Genève, Switzerland}

\newcommand{\affiliationAnthony}{Now at: Université Côte d’Azur, CNRS, Institut de Physique de Nice, Nice, France.}

\DeclareSIUnit\permille{\text{\textperthousand}}
\DeclareSIUnit{\belmilliwatt}{Bm}
\DeclareSIUnit{\dBm}{\deci\belmilliwatt}
\DeclareSIUnit{\bps}{bps}
\DeclareSIUnit{\bit}{b}

\graphicspath{{./Figures/}}

\begin{document}
\title{Modulator-free, self-testing quantum random number generator} 

\author{Ana Blázquez-Coído}\email{ablazquez@vqcc.uvigo.es}
\affiliation{\affiliationUVigo}
\affiliation{\affiliationComGroup}
\affiliation{\affiliationAtlanttic}
\author{Fadri Grünenfelder}
\affiliation{\affiliationUVigo}
\affiliation{\affiliationComGroup}
\affiliation{\affiliationAtlanttic}
\author{Anthony Martin}
\affiliation{\affiliationGeneva}
\affiliation{\affiliationAnthony}
\author{Raphael Houlmann}
\affiliation{\affiliationGeneva}
\author{Hugo Zbinden}
\affiliation{\affiliationUVigo}
\affiliation{\affiliationComGroup}
\affiliation{\affiliationAtlanttic}
\affiliation{\affiliationGeneva}
\author{Davide Rusca}
\affiliation{\affiliationUVigo}
\affiliation{\affiliationComGroup}
\affiliation{\affiliationAtlanttic}

\begin{abstract}
Quantum random number generators (QRNGs) use the inherent unpredictability of quantum mechanics to generate true randomness, as opposed to classical random number generators. However, ensuring the authenticity of this randomness still requires robust verification. Self-testing QRNGs address this need by enabling the validation of the randomness produced based on the observed data from the experiment while requiring few assumptions. In this work, we present a practical, self-testing QRNG designed to operate with an untrusted measurement device and a partially characterized source, allowing the user to check the adequate functioning of the setup in real time. Our experiment yields a rate of certified random bits of 450kbps.
\end{abstract}

\maketitle

\section{Introduction}

Recently, quantum random number generators (QRNGs) have obtained notable attention and undergone substantial development \cite{Herrero2017, Ma2016, Mannalath2023}, driven by their utility across diverse fields. They are particularly crucial in cryptography, where truly unpredictable and private randomness is essential. These devices offer a compelling advantage compared to classical random number generators: the inherent randomness of their output strings, derived from the principles of quantum mechanics. Despite the proliferation of a vast range of QRNG designs, the key challenge lies in validating the quantum origin of the randomness produced \cite{van2017,Michel2019,gehring2020}. This difficulty in discerning the quantum effects from noise generated by uncharacterized sources gives way to two different approaches to the design of QRNGs: device dependent and device independent. The device dependent approach requires comprehensive characterization of the device to discern and mitigate potential sources of classical noise that could be exploited by malicious parties.
Alternatively, the device independent approach \cite{Colbeck2009,Pironio2010,Colbeck2011} offers a different paradigm, enabling the certification of output randomness even in scenarios where the device itself is partially constructed by an adversary. 
This approach promises the highest level of security, but requires to perform experimentally challenging loophole-free Bell tests. Thus, the complexity of DI-QRNG devices and their significantly lower rates pose substantial hurdles to widespread adoptions \cite{liuyang2018}. 

In the middle ground, we can find different approaches such as source-independent QRNGs \cite{zhu2016}, measurement-device-independent (MDI) QRNGs \cite{Cao2015} and semi-device-independent or self-testing QRNGs \cite{lunghi2015,brask2017,rusca2019,avesani2021}. The goal of these approaches is to design QRNGs that are easier to implement than DI-QRNGs but still maintain robust security measures. While source-independent QRNGs work with an untrusted source but require characterization and trust on the measurement devices, MDI QRNGs require trust only in the source while being robust to adversarial manipulation of the measurement devices. Semi-device independent QRNGs balance between full independence and dependence by introducing specific, verifiable assumptions about the device to maintain security without full trust in any component. This approach aims to develop QRNGs with a high level of security while minimizing assumptions about the underlying device. Recent studies have introduced various approaches within this framework, leveraging assumptions about the dimensionality \cite{lunghi2015}, overlap \cite{brask2017}, or energy \cite{rusca2019,avesani2021} of the quantum states produced by the device. These advancements represent significant strides towards achieving a balance between simplicity and security in QRNG design.
In this work we present a simple, self-testing QRNG that operates with an untrusted measurement device and a partially characterized source based on a prepare-and-measure framework. This scheme obviates the need for costly electro-optical modulators and their associated driving electronics, as required in previous works \cite{Rusca2020}. Consequently, the implementation cost of the setup is substantially diminished.
\section{Scheme}

\begin{figure}
	\includegraphics[width = 1\columnwidth]{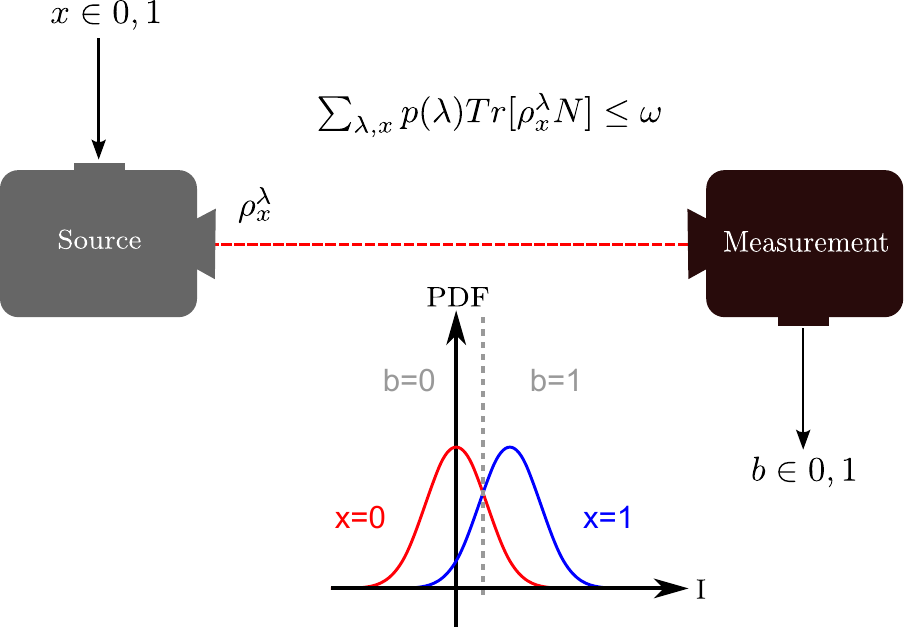}
	\caption{\label{fig:states} Scheme of the prepare-and-measure scenario considered, assuming that the states sent have a bounded energy. Also represented is the probability density function of the current output of a balance homodyne detector when the states $\ket{0}$ (x=0) and  $\ket{\alpha}$ (x=1) are measured. The determination of the value of $b$, the binary output received from the measurement device, is done by setting a threshold that separates the two possible outcomes (represented by a grey dashed line).}
\end{figure}

We consider a prepare-and-measure scenario where the preparation device takes a binary input $x$ and the measurement device outputs the binary value $b$. The preparation device sends a weak coherent state ($\ket{\alpha}$) when $x$=1 and a vacuum state ($\ket{0}$) when $x$=0. 
The measurement device employs homodyne detection to distinguish between these states, and the output value is chosen when the detector current is below ($b$=0) or above ($b$=1) a certain threshold, represented in \autoref{fig:states}. The measurement part of the experiment is considered untrusted while the source part has some energy limitation in the states it can produce. We also assume that no entanglement is shared between the state preparation and the measurement parts, so no information is leaked about the input $x$. The probability of the output $b$ conditioned on the input $x$ can be defined as:

\begin{equation}
    p(b\vert x) = \sum_{\lambda}p(\lambda)Tr[\rho^{\lambda}_x M^{\lambda}_b]
\end{equation}
 
where $\rho^\lambda_x$ is the quantum state that the source prepares, $M^\lambda_b$ is the positive-operator valued measure (POVM) element that corresponds to the output $b$ and $\lambda$ is an arbitrary classical variable that is meant to represent the classical noise. In order to certify the quantum origin of the randomness generated by output $b$, we need to track these correlations between input and output $p(b\vert x)$ and the average energy per pulse must respect an upper bound \cite{van2017,van2019,rusca2019}. This upper bound is described in the following equation:

\begin{equation}
    \sum_{\lambda,x}p(\lambda)Tr[\rho^{\lambda}_x N] \leq \omega
\end{equation}

with $N$ being the photon number operator and $\omega$ the chosen bound of the average photon number of the pulses sent by the preparation device, also referred to as energy bound. With the conditional probabilities $p(b\vert x)$ and semidefinite programming (SDP) we can get a lower bound on the conditional Shannon entropy $H(B\vert X, \Lambda)$ \cite{van2017}. We use the worst-case conditional smooth min-entropy:
\begin{equation}
    H^{\epsilon'}_{min}(B\vert X, \Lambda) \geq n \left( h - c\sqrt{\frac{log(\epsilon/2)}{n}}-d\frac{log(\epsilon/2)}{n}\right)
\end{equation}

this represents the bits that a strong extractor can output from the raw string of bits $B$ generated by the experiment. Using this, we can estimate the information that Eve has about the random bits $b$ and extract truly random bits $b'$ (See Ref.\cite{rusca2019,van2019} for more details).  

\section{Implementation}

\begin{figure}
	\includegraphics[width = 1\columnwidth]{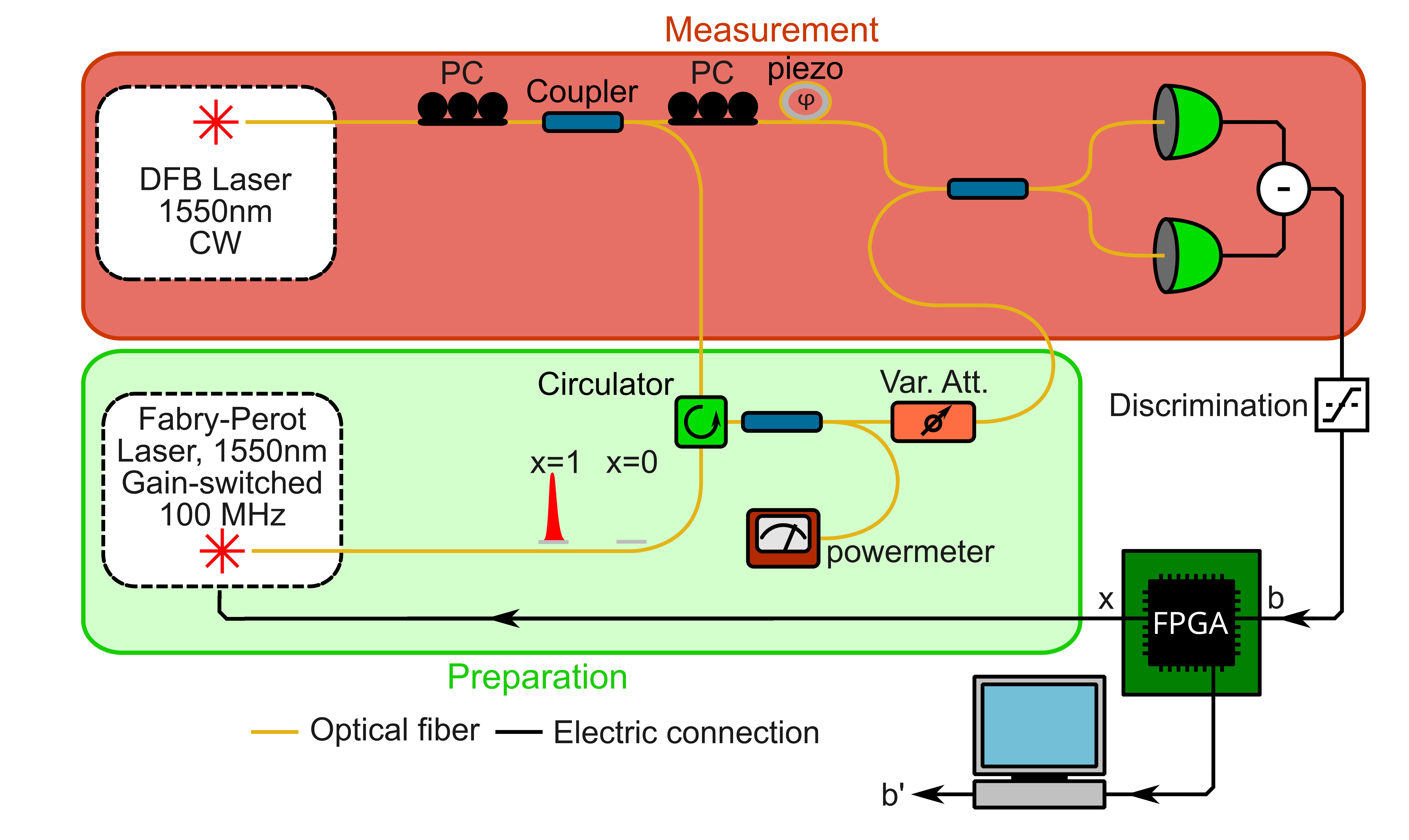}
	\caption{\label{fig:setup_qrng} Experimental setup of the QRNG, divided in trusted and partially characterized (green box) and untrusted (red box). The top arm corresponds to the local oscillator and the bottom part is where the state preparation is performed. 
	}
\end{figure}

\begin{figure}
	\includegraphics[width = 1\columnwidth]{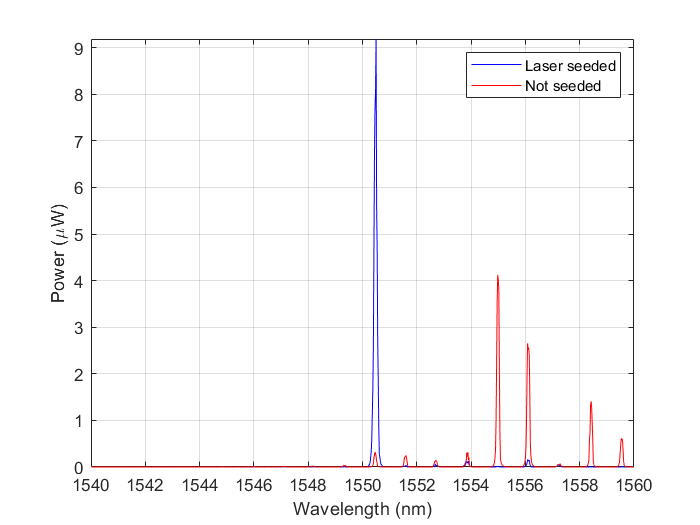}
	\caption{\label{fig:seed} Comparison of the spectral analysis of the Fabry-Perot laser while seeded (shown in blue) and not seeded (red).}
\end{figure}

The setup can be divided in two parts: preparation, which is partially trusted, and measurement, which is completely untrusted (see \autoref{fig:setup_qrng}). The preparation part mainly consists on a pulsed Fabry-Perot (FP) laser (Koheron LD101) that operates in the telecommunication wavelength (1550nm) and is modulated at a repetition rate of 12.5MHZ. The pulses coming from the laser are then attenuated by a variable attenuator to the single photon level. Before the attenuator, a power meter is used to keep track of the average power output from the FP laser, so that we can set the correct attenuation to achieve the target photon number per pulse. This measurement includes any possible noise from the seeding process (described below), including both direct leakage through the circulator (port 1 to port 3) and backreflections from port 2 to port 3. The upper bound of the mean photon number per pulse is calculated using the following formula:

\begin{equation}
 \bar{\mu} = \frac{P_{in}\cdot \eta \cdot\Delta{t}}{h\cdot \nu_{min} \cdot f_{rep}}
\end{equation}

where $P_{in}$ is the power measured with the power meter, $\eta= P_{in}/P_{out}$ ($P_{out}$ being the power existing the preparation stage),  $\Delta{t}$ is the modulation duration, $h$ is the Planck constant, $\nu_{min}$ is the lower bound on the frequency of the laser and $f_{rep}$ is the repetition rate. The homodyne measurement arrangement, consists of a  distributed feedback (DFB) laser (Gooch and Housego AA0701) that acts as a local oscillator for the balanced photodectector (Thorlabs PDB480C-AC). As this part is left completely untrusted, it doesn't have to be characterized at all. The DFB laser is also used to seed the FP laser via injection locking, to control the phase coherence between the two lasers \cite{paraiso2019}. This is because phase coherence is necessary for the homodyne detection to yield meaningful results. The locking is performed by adjusting the temperature of the DFB laser with a temperature controller while checking the spectrum of the FP laser with a spectrum analyzer. When the laser is correctly seeded, all modes except one disappear. A comparison between the outputs of the FP laser when seeded and not seeded can be seen in \autoref{fig:seed}. The extinction ratio of the seeded laser was measured as the ratio of the optical power in the seeded mode with respect to the power in the unseeded modes, giving a result of -13.47dB. 

In order to maintain phase stability within the interferometer, we have set a piezoelectric cylinder connected to a feedback loop in one of it's arms, around which approximately half a meter of optical fiber is wound. The feedback loop provides the conditional probabilities, allowing for phase optimization using the piezo to enhance entropy values.  We employ two polarization controllers within the setup: one preceding the interferometer, and another situated in the local oscillator arm (see \autoref{fig:setup_qrng}). The first is used to optimize  the seeding of the FP laser, while the latter ensures alignment with the polarization of the opposing arm.

We implement an on-off keying (OOK) scheme where the FP laser, depending on the binary input $x$ it receives, produces a coherent or a vacuum state. The states sent by the FP laser are distinguished by the homodyne detection via a quadrature measurement, which outputs an electrical signal that is then sent to the discrimination card in order to produce a binary output $b$. This output is selected by a threshold level that is manually set in order to get equal values on the conditional probabilities. A field-programmable gate array (FPGA) controls the laser diode and registers detection outcomes $b$ and forwards them to a computer, while also keeping track of the conditional probabilities p(b$\vert$x). The computer then calculates the min-entropy by feeding these conditional probabilities and the bound on the average energy to the SDP.




\section{Results}

\begin{figure}
	\includegraphics[width = 1\columnwidth]{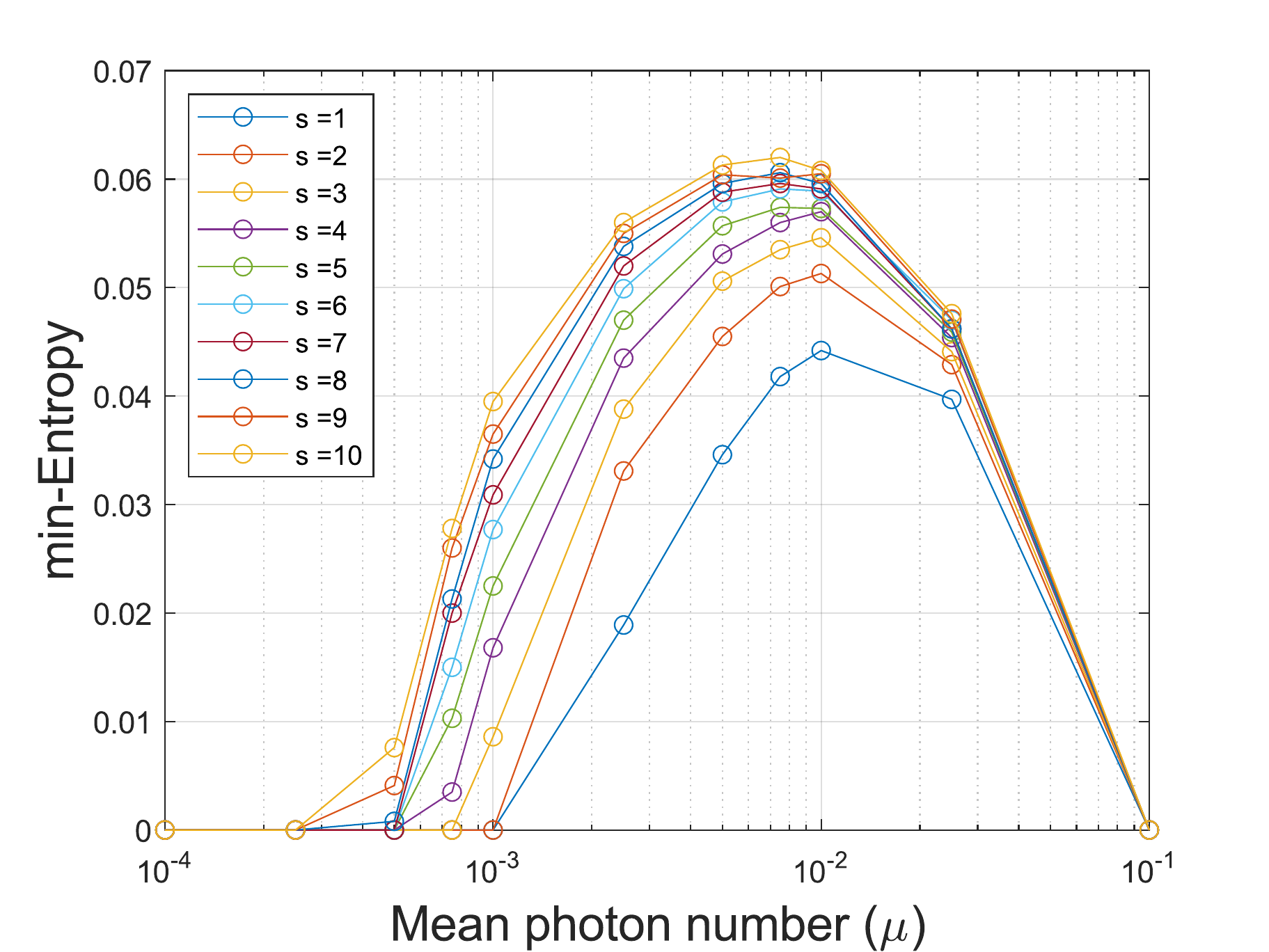}
	\caption{\label{fig:entropyvsmu} Maximum value of the min-entropy achieved with respect to the mean photon number used. This value was measured for block sizes from one to ten seconds (s).}
\end{figure}

First, we simulate the dependence of maximum extractable randomness with respect to the chosen energy bound (expressed in the mean photon number of the prepared pulses). We assume different acquisition times, from one to ten seconds. As shown in \autoref{fig:entropyvsmu} the amount of extractable randomness has a maximum at around $10^{-2}$ photon per pulse. The results obtained show worse performance in entropy per pulse when compared to previous implementations \cite{Rusca2020}. However, this trade-off enables a simpler and more cost-effective setup.


\begin{figure}
	\includegraphics[width = 1\columnwidth]{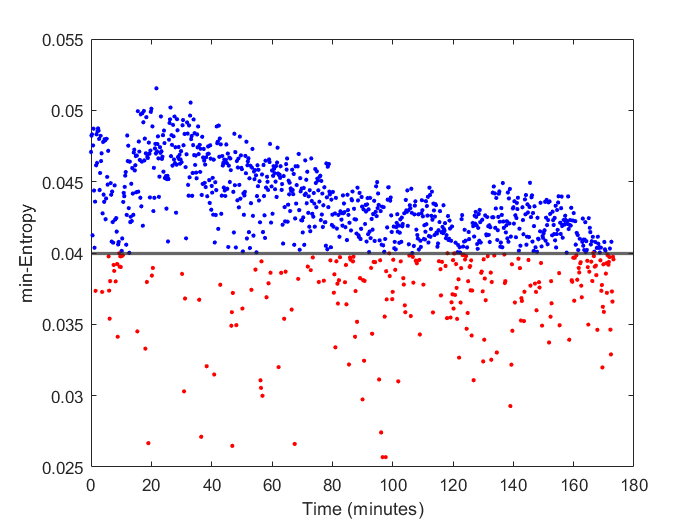}
	\caption{\label{fig:stability} Stability of the system measured for around three hours. Each point corresponds to a 10s measurement.}
\end{figure}

\begin{figure}
	\includegraphics[width = 1\columnwidth]{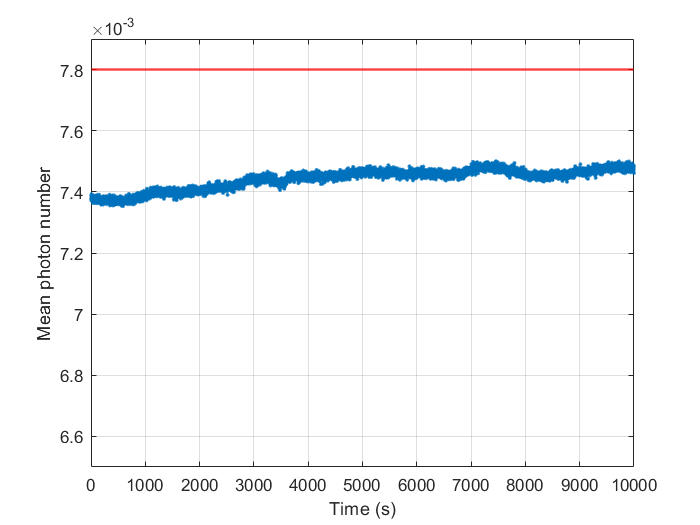}
	\caption{\label{fig:photonlimit} Mean photon number measured in the experiment as a function of time, with every point being a measurement over one second. The red line represents the energy bound, which is never violated.}
\end{figure}

We also measured the stability of the system for around 3 hours (see \autoref{fig:stability}), where it can be seen that the maximum extractable entropy stays stable for at least one hour thanks to the use of temperature controllers and the piezoelectric cylinder. The stability of the setup could be fixed by adding a feedback loop to the temperature controller of the seeding laser and using polarization maintaining fiber so that both the seeding and the polarization stay stable for longer periods of time. We can also note in \autoref{fig:photonlimit} that during the entire duration of the measurement the energy bound was never violated, as the mean photon number stayed below the limit. By setting a threshold of the min-entropy at 0.04 we can see that 80\% of the points exceed this value. The aforementioned threshold has to be set before running the experiment, according to the theory \cite{van2017}. With this, we get a rate of certified randomness of 450kbps during the first hour, which could increase with faster modulation of the pulsed laser.

\section{Discussion}

\begin{figure}
	\includegraphics[width = 1\columnwidth]{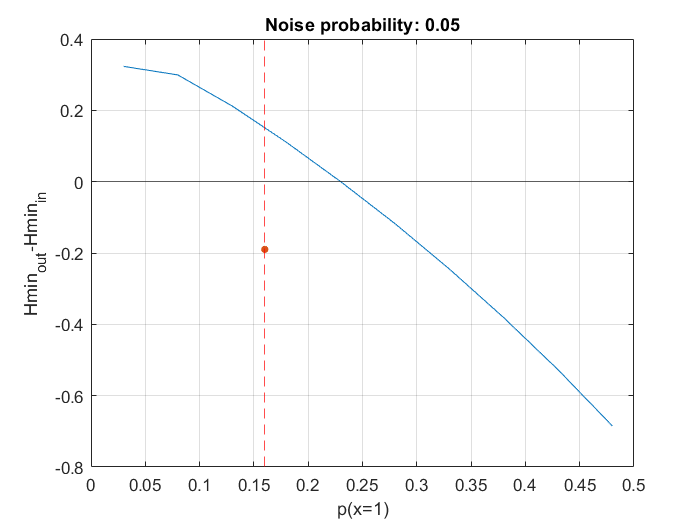}
	\caption{\label{fig:hmininout} Simulation results of the generation of min-entropy with respect to the probability to send x=1 using 5\% noise probability. Red dotted line represents the probability used in the experiment, with the red dot representing the experimentally obtained difference.}
\end{figure}

For the correct functioning of the QRNG, the input $x$ must be independent from the output $b$. The way the experiment was realized, the input $x$ was generated using a pseudo random number generator \cite{brask2017}, but it could also be obtained, for example, from a source of public randomness. This public randomness would be distillated into private randomness by the experiment, making it secure for cryptographic purposes. Another way the randomness of the input $x$ could be generated is by using the entropy that our device itself outputs, but it would require that the devices outputs more entropy than we need to provide in the input.
For this reason, we simulated the difference in the min-entropy that is provided to the system via the probability of the input variable $x$ and the min-entropy that we can get out of it, shown in \autoref{fig:hmininout}. We used a lower bound for the noise probability of 5\%, which is the minimum that we can get with this system, as per the clearance of the homodyne detector. With the probability to send $x=1$ used in the experiment being 0.16, the simulation results indicate that we could be outputting more entropy than we insert in but we would need to reduce the noise introduced by other components into the experiment. Since our scheme has a low probability of sending a pulse with respect to sending vacuum, the results remain largely unaffected by the patterning effects of the pulsed laser. This means we could modulate at higher frequencies without needing to change any component. In contrast, using an electro-optical modulator would require a broadband device, which is significantly more expensive. This price difference is substantial, as such modulators cost thousands of euros, whereas a standard diode laser is widely available and can be purchased for just tens of euros. This allows for cost-efficient scalability of the setup with respect to modulation frequency.

\section{Conclusion}

This semi-DI QRNG eliminates the need for costly electro-optical modulators used in previous works \cite{Rusca2020} and their associated driving electronics, significantly reducing implementation costs. Furthermore, we achieve an extraction rate of certified quantum randomness of around 450kbps. There is still room for further refinement, particularly in ensuring long-term stability to maintain high entropy levels over extended durations, which could be improved with a slave laser that is easier to stabilize in temperature and the use of polarization maintaining fiber.\\



\textbf{Abbreviations:} QRNG quantum random number generator; DI-QRNG device independent QRNG; MDI measurement device independent; POVM positive-operator value measurement; SDP semidefinite programming; FP Fabry-Perot; DFB distributed feedback; OOK On-Off-Keying; FPGA field-programmable gate array.\\

\section{Declarations}

\textbf{Avaliability of data and materials:} The data is available upon reasonable request.\\

\textbf{Competing interests:} The authors declare no competing interests.\\

\textbf{Funding:} This work was supported by the Galician Regional Government (consolidation of Research Units: AtlantTIC), MICIN with funding from the European Union NextGenerationEU (PRTR-C17.I1) and the Galician Regional Government with own funding through the "Planes Complementarios de I+D+I con las Comunidades Autónomas" in Quantum Communication and the European Union’s Horizon Europe Framework Programme under the project "Quantum Security Networks Partnership" (QSNP, grant agreement No101114043). It was also supported by the Swiss State Secretariat for Research and Innovation (SERI) (UeM019-3).\\

\textbf{Authors' contributions:} AM conceived the original idea. DR, FG and HZ designed the optical setup. AB and FG built the experimental setup and took data from the experiment. RH programmed the FPGA used in the setup. AB analyzed the data and wrote the first version of the manuscript. DR and HZ supervised the work. All authors reviewed and approved the final manuscript.\\

\textbf{Acknowledgements}: We thank Thomas Van Himbeeck for his work in the simulation code.\\

~


\bibliography{bibliography}
\end{document}